\documentclass[lettersize,journal]{IEEEtran}
\usepackage{amsmath,amsfonts}
\usepackage{algorithmic}
\usepackage{algorithm}
\usepackage{array}
\usepackage{textcomp}
\usepackage{stfloats}
\usepackage{url}
\usepackage{verbatim}
\usepackage{graphicx}
\usepackage{subcaption}
\usepackage{xcolor}
\usepackage{cite}
\usepackage{balance}
\usepackage{makecell}
\usepackage{amssymb} 
\usepackage{amsmath}

\newcommand{\mat}[1]{\mathbf{#1}}

\newcommand{\sgn}{\mathop{\mathrm{sgn}}}

\begin{document}

\title{Scalable Digital Compute-in-Memory Ising Machines for Robustness Verification of Binary Neural Networks}

\author{\IEEEauthorblockN{
Madhav Vadlamani\textsuperscript{*, 1},
Rahul Singh\textsuperscript{*, 2},
Yuyao Kong\textsuperscript{1},
Zheng Zhang\textsuperscript{$\dagger$, 2},
Shimeng~Yu\textsuperscript{$\dagger$,1}
} \\
\IEEEauthorblockA{\textsuperscript{1}School of Electrical and Computer Engineering,
Georgia Institute of Technology, Atlanta, GA, 30332} \\
\IEEEauthorblockA{\textsuperscript{2}Department of Electrical and
Computer Engineering, University of California, Santa Barbara, CA, 93106} \\
\IEEEauthorblockA{\textsuperscript{*}Equal contribution.}
\IEEEauthorblockA{$^{\dagger}$Corresponding authors: Zheng Zhang (zhengzhang@ece.ucsb.edu), and Shimeng Yu (shimeng.yu@ece.gatech.edu).}
}

\markboth{Journal of \LaTeX\ Class Files,~Vol.~XX, No.~XX, AUGUST ~2026}%
{Shell \MakeLowercase{\textit{et al.}}: A Sample Article Using IEEEtran.cls for IEEE Journals}

\maketitle

\begin{abstract}
Verification of binary neural network (BNN) robustness is NP-hard, as it can be formulated as a combinatorial search for an adversarial perturbation that induces misclassification. Exact verification methods therefore scale poorly with problem dimension, motivating the use of hardware-accelerated heuristics and unconventional computing platforms, such as Ising solvers, that can efficiently explore complex energy landscapes and discover high-quality solutions. In this work, we reformulate BNN robustness verification as a quadratic unconstrained binary optimization (QUBO) problem and solve it using a digital compute-in-memory (DCIM) SRAM-based Ising machine. Instead of requiring globally optimal solutions, we exploit imperfect solutions produced by the DCIM Ising machine to extract adversarial perturbations and thereby demonstrate the non-robustness of the BNN. The proposed architecture stores quantized QUBO coefficients in approximately 9.1~Mb of SRAM and performs annealing in memory via voltage-controlled pseudo-read dynamics, enabling iterative updates with minimal data movement. Experimental projections indicate that the proposed approach achieves a $178\times$ acceleration in convergence rate and a $1538\times$ improvement in power efficiency relative to conventional CPU-based implementations. 
\end{abstract}


\section{Introduction}

Recent advances in deep learning have driven substantial growth in the computational and energy cost of both training and inference, motivating network designs that can be deployed efficiently on resource-constrained platforms. Binary neural networks (BNNs) are a compelling option in this regime. By constraining weights, biases, and/or activations to binary values, inference can be implemented largely with bitwise primitives (e.g., XNOR/XOR) and low-cost accumulation, substantially reducing hardware complexity and energy~\cite{Blott2018,Zhu2020}. Despite these efficiency benefits, neural networks—including BNNs—are known to exhibit brittleness to small, structured input perturbations~\cite{galloway2018attacking,Szegedy2014,Goodfellow2015}. Minor modifications to the input can induce incorrect outputs with significant implications for security, decision-making, and health-related applications~\cite{QIN2020107281,rastegari2016xnornetimagenetclassificationusing}. 

Accordingly, robustness verification, which certifies whether a model’s output is provably invariant within a specified perturbation set, has become a critical step for trustworthy deployment. Existing verification approaches are predominantly executed on classical computing platforms and include (i) sound but often conservative over-approximation methods based on abstract interpretation and bound propagation, and (ii) exact constraint-solving approaches (e.g., SMT/MILP-based verification) that provide formal guarantees but can scale poorly to large networks~\cite{Singh2019,Wang2021,Xiang2018,Gehr2018,Weng2018,Weng2019,Huang2020,Fazlyab2019}. In contrast, statistical or heuristic testing-based techniques are generally incomplete and do not provide worst-case robustness guarantees. 

For BNNs in particular, the discrete structure introduced by binarized weights and activations further complicates verification, and existing methods remain predominantly classical and are typically formulated as combinatorial optimization problems~\cite{Narodytska2018,Narodytska2020}. Such formulations often produce highly nonconvex and rugged optimization landscapes with numerous local minima, making the search for optimal solutions computationally demanding and, in general, NP-hard when addressed using conventional computing methods. These characteristics motivate the use of Ising-based and quantum-inspired hardware platforms that naturally solve combinatorial optimization problems by exploiting physical energy minimization. These unconventional computing platforms provide a promising alternative, as they can leverage massive parallelism, collective dynamics, and stochastic or coherent evolution to more effectively explore complex energy landscapes than conventional sequential solvers. 

\begin{figure*}[t]
    \begin{center}
        \centerline{\includegraphics[width=\textwidth]{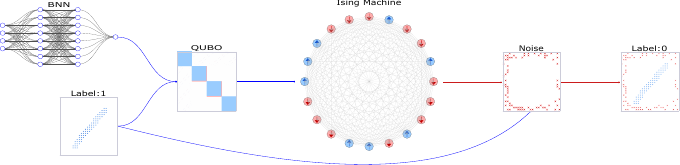}} 
        \caption{Flowchart of the proposed robustness-verification pipeline. A QUBO instance is constructed from a pre-trained BNN and a correctly classified input-ouput pair. An Ising machine then searches for a perturbation (noise) that modifies the input and induces a change in the BNN output, thereby yielding an adversarial example.}
        \label{robustness}
    \end{center} 
\end{figure*}

In~\cite{singh2026}, the BNN verification problem is converted to a QUBO instance which can be solved with Ising machines. This approach, in principle, be extended to larger BNNs and more diverse binarized datasets. However, the resulting QUBO matrices are often densely connected, which can quickly stress coupling storage, bandwidth, and update parallelism in scalable annealers. These characteristics motivate the development of hardware accelerators that can efficiently handle large state spaces and dense couplings. In our prior work~\cite{10.1145/3649329.3655673}, we addressed several of these bottlenecks while solving the Traveling Salesman Problem (TSP)~\cite{Hoffman2013, 10.1145/3649329.3655673, Kong2026Compact} using an SRAM-based DCIM Ising annealer. The proposed architecture combines hierarchical clustering to reduce the effective problem density, compact weight mapping that exploits sparsity through partial-column accumulation, and chromatic Gibbs sampling to update conditionally independent clusters in alternating phases. We further validated this approach experimentally through energy-efficient, parallel annealing on a compact DCIM macro prototype chip fabricated in a 28\,nm CMOS process~\cite{Kong2026Compact}.

Another critical limitation in many SRAM-based annealers is the overhead of dedicated digital pseudo-random number generators (e.g., LFSRs) required to supply stochasticity for sampling-based updates \cite{10454294}. Our work instead leverages a fundamental architectural innovation \cite{10.1145/3649329.3655673}: we inject noise into the weight matrix rather than the spin states, converting otherwise static spatial mismatch into effective temporal randomness through clock-cycled device utilization. Concretely, by operating SRAM-based DCIM under controlled conditions (e.g., pseudo-read/reduced-$V_{\mathrm{DD}}$ modes), intrinsic bitcell variability—experimentally characterized in prior silicon \cite{7350099} - is harnessed as a native entropy source that perturbs the coupling terms during annealing. This design eliminates the need for external RNG hardware while preserving annealing efficacy, and provides a scalable, circuit-native mechanism for stochastic Ising optimization that naturally extends beyond TSP to constraint-rich verification tasks such as robustness verification of BNNs.

To the best of our knowledge, this work demonstrates for the first time an imperfect-solution verification workflow on an SRAM-based DCIM Ising architecture, establishing an end-to-end hardware path from QUBO-encoded verification constraints to annealing-based adversarial example generation. The proposed formulation and workflow are compatible with a broad solver ecosystem, including Ising machines, quantum annealers, and quantum-inspired optimization platforms. Figure~\ref{robustness} illustrates the proposed robustness-verification pipeline. A QUBO instance is constructed from a pre-trained BNN and a correctly classified input–output pair. The Ising machine then explores the corresponding energy landscape to generate candidate perturbations, which are subsequently validated through forward inference to identify adversarial examples.

The remainder of this paper is organized as follows. Section~II summarizes the QUBO formulation for BNN robustness verification used in this work, adapted from prior work~\cite{singh2026}. Section~III describes the proposed SRAM-based DCIM Ising annealing architecture and its control flow. Section~IV presents simulation results, including solution quality and end-to-end adversarial perturbation recovery using the proposed imperfect-solution verification paradigm, which leverages imperfect (non-globally optimal and potentially constraint-violating) solutions produced by Ising solvers. In contrast to approaches that require fully constraint-satisfying global minima, which are often intractable for large instances, we show that suboptimal samples can still encode valid adversarial examples when validated by forward inference on the BNN, providing a useful intermediate regime between exact verification and heuristic search. Finally, Section~V concludes and discusses implications for scalable hardware verification.

\section{Background : Robustness Verification Problem and its QUBO Formulation}

In this section, we briefly explain how a BNN robustness verification problem can be converted into a QUBO instance~\cite{singh2026}. Robustness verification for $\mathrm{BNN}(\cdot)$, given an input $\mat{x}$ and its associated label $y$, seeks to determine whether the network's prediction remains invariant under all admissible perturbations $\boldsymbol{\tau} \in \boldsymbol{\Omega}$. The network is said to be robust at $\mat{x}$ if
\begin{equation}
    y = \mathrm{BNN}(\mat{x} + \boldsymbol{\tau}), \quad \forall \boldsymbol{\tau} \in \boldsymbol{\Omega}.
\end{equation}
Conversely, the network is non-robust if there exists a perturbation $\boldsymbol{\tau} \in \boldsymbol{\Omega}$ such that
\begin{equation}
    \exists\, \boldsymbol{\tau} \in \boldsymbol{\Omega}
    \;\; \text{s.t.} \;\;
    \mathrm{BNN}(\mat{x} + \boldsymbol{\tau}) \neq y.
\end{equation}

Ising hardware requires the target task to be expressed as an Ising Hamiltonian. However, a direct, general mapping from BNN robustness verification to an Ising form is typically unavailable. The framework in~\cite{singh2026} therefore proceeds in two stages: first, it casts robustness verification as an optimization problem that searches for an adversarial perturbation capable of inducing misclassification; it then converts this objective into an equivalent QUBO/Ising Hamiltonian for execution on the Ising machine.

Let $\mat{x}$ denote the nominal input and $\mat{x}'$ an adversarially perturbed input. The perturbation magnitude is measured via a distance
\begin{equation}
    d(\mat{x},\mat{x}') = \left\|\mat{x}-\mat{x}'\right\|,
\end{equation}
which, in the binary setting considered here, corresponds to the number of flipped bits. For a maximum allowable perturbation budget $\epsilon$, robustness verification can be written as
\begin{equation}
    \begin{aligned}
        \underset{\boldsymbol{\tau}}{\min} \quad & d(\mat{x}, \mat{x}') \\
        \text{s.t.} \quad & d(\mat{x}, \mat{x}') \leq \epsilon, \\
                          & \mathrm{BNN}(\mat{x}) \neq \mathrm{BNN}(\mat{x}'), \\
                          & \mat{x}' = \mat{x} + \boldsymbol{\tau}.
    \end{aligned}
    \label{Eq:MinimizationProblem}
\end{equation}

The optimization problem in~\eqref{Eq:MinimizationProblem} is reformulated as a quadratic constrained Boolean optimization (QCBO) instance by expressing the BNN inference constraints and the adversarial constraints directly in the Boolean domain. This requires rewriting arithmetic operations in terms of Boolean primitives or enforcing them via Boolean variable constraints. In particular, the element-wise multiplications in binary-weight inference are realized using XNOR operations, and the layer nonlinearity is enforced by extracting the sign of an accumulated sum of XNOR outcomes. The perturbed input is obtained by an element-wise XOR between the nominal input $\mat{x}$ and the perturbation vector $\boldsymbol{\tau}$. 

The next step is to convert the QCBO formulation into a QUBO, which serves as a standard input format for many Ising-based hardware platforms. To this end, the QCBO constraints are incorporated into a single unconstrained quadratic objective using the standard penalty method~\cite{Sasdelli2021, singh2026}. If the resulting polynomial contains terms of degree higher than two, its order is reduced through quadratization by introducing auxiliary Boolean variables. The resulting objective is a standard QUBO and can be written in matrix form as
\begin{equation}
    H(\mathbf{q}) = \mathbf{q}^{T}\mathbf{Q}\mathbf{q},
    \label{eq:qubo_matrix_form}
\end{equation}
where $\mathbf{q}$ denotes the Boolean vector over which the optimization is performed, including perturbations, perturbed inputs, intermediate layer outputs, and auxiliary components introduced during quadratization. The matrix $\mathbf{Q}$ represents the corresponding QUBO coefficient matrix, an example of which is illustrated in Fig.~\ref{robustness}.

A perfect solution corresponds to a configuration in which the perturbation vector $\boldsymbol{\tau}$ induces a misclassification while all auxiliary variables remain consistent with the modeled BNN inference and verification constraints. If such a perturbation exists that simultaneously produces a valid adversarial example within the allowed budget and satisfies all encoded BNN computation constraints, the corresponding configuration represents the global minimum of the QUBO objective. In practice, the full QUBO generated by a realistic multi-layer BNN and high-dimensional inputs can be prohibitively large, motivating scalable solver architectures; this motivates the SRAM-based DCIM Ising solver described in the next section, which is designed to efficiently search for low-energy solutions under such constraint-rich formulations.

\section{The SRAM DCIM Ising Architecture}

\subsection{Ising Formulation and QUBO Mapping}

Consider the Ising Hamiltonian
\begin{equation}
H(\boldsymbol{\sigma}) \;=\; -\sum_{i<j} J_{ij}\sigma_i\sigma_j \;-\; \sum_i h_i\sigma_i \;+\; C,
\label{eq:ising}
\end{equation}
where $\sigma_i \in \{-1,+1\}$ denotes the spin of site $i$, $J_{ij}$ are pairwise couplings, $h_i$ are local fields, and $C$ is an additive constant.

To express \eqref{eq:ising} as a QUBO, we introduce binary variables $q_i \in \{0,1\}$ through the standard affine transformation
\begin{equation}
q_i \;=\; \frac{1+\sigma_i}{2}
\qquad\Longleftrightarrow\qquad
\sigma_i \;=\; 2q_i - 1 .
\label{eq:bit_spin}
\end{equation}

Substituting \eqref{eq:bit_spin} into \eqref{eq:ising} and collecting terms yields a binary quadratic form equivalent to \eqref{eq:qubo_matrix_form}, where the resulting QUBO matrix has off-diagonal entries proportional to the Ising couplings (capturing the pairwise interaction terms) and diagonal entries that aggregate the local-field contributions together with the coupling-dependent bias induced by the bit–spin transformation, while all terms independent of $\mathbf{q}$ are absorbed into the constant offset.

\begin{figure*}[hbtp]
    \begin{center}                    
        \centerline{\includegraphics[width=0.95\textwidth]{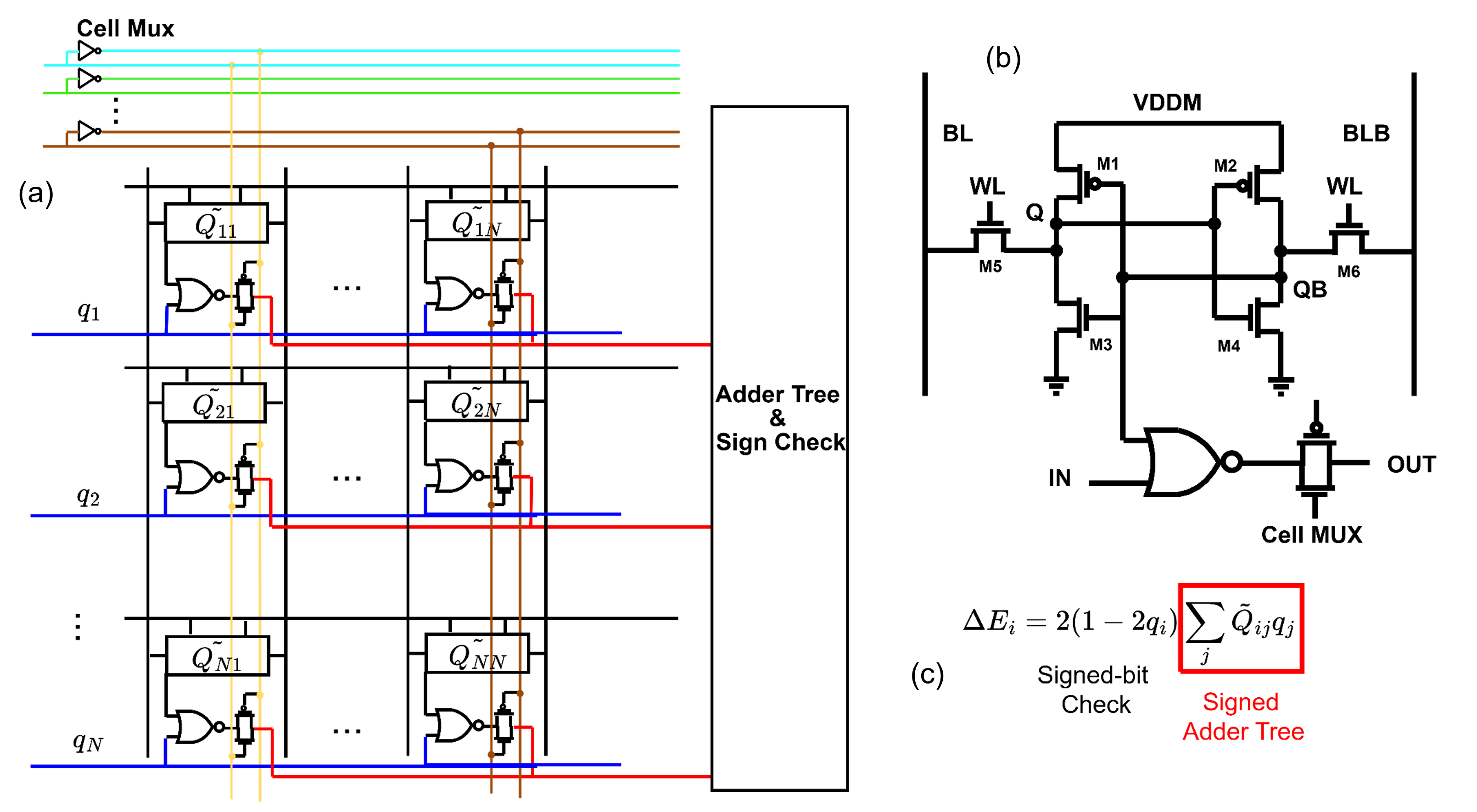}} 
        \caption{DCIM-based Hamiltonian update engine. (a) SRAM array storing quantized weights $\tilde{Q}_{ij}$, with spin inputs $q_j$ broadcast on wordlines and column signals reduced by a signed adder tree. (b) A 6T SRAM bitcell with a per-cell MUX enabling selective weight participation during MAC and pseudo-read noise injection via $V_{\mathrm{DDM}}$ scaling. (c) In-memory per-spin update computation, where the signed adder tree forms $\sum_j \tilde{Q}_{ij} q_j$ and a sign check implements the flip decision based on the sign of $\Delta E_i$.}
        \label{sram_array}
    \end{center} 
\end{figure*}

The spin-update decision is based on the local energy change incurred by flipping a single spin while holding the remaining spins fixed. For site $i$, define
\begin{equation}
\Delta E_i \;\triangleq\; H(\sigma_1,\ldots,-\sigma_i,\ldots,\sigma_N)\;-\;H(\sigma_1,\ldots,\sigma_i,\ldots,\sigma_N).
\end{equation}
Using \eqref{eq:ising}, this per-spin energy difference simplifies to
\begin{equation}
\Delta E_i \;=\; 2\,\sigma_i\!\left(\sum_{j\neq i} J_{ij}\sigma_j \;+\; h_i\right),
\label{eq:deltaE_ising}
\end{equation}
i.e., the flip cost is determined by the product of the current spin $\sigma_i$ and its effective local field given by the weighted sum of neighboring spins plus the local-field term.

Using \eqref{eq:bit_spin}, an equivalent update rule can be written directly in the QUBO domain. Specifically, for the quadratic objective in \eqref{eq:qubo_matrix_form}, flipping a single variable $q_i \leftarrow 1-q_i$ while holding all other variables fixed changes the objective by
\begin{equation}
\Delta E_i \;=\; (1-2q_i)\!\left(Q_{ii} \;+\; 2\sum_{j\neq i} Q_{ij}q_j\right),
\label{eq:deltaE_qubo}
\end{equation}
where the term in parentheses is the effective local field acting on $q_i$ under the QUBO parametrization. This expression is the direct analogue of \eqref{eq:deltaE_ising} and forms the basis of the single-site update used in annealing for QUBO objectives.

\subsection{Weight Storage and Quantization}

In the proposed DCIM architecture, the matrix entries are loaded into the SRAM array and the spins are stored separately in a register that drives the array inputs during iterative updates. Each coupling coefficient is first quantized to a signed fixed-point representation (8 bits in this work) and then decomposed into bit-slices that are mapped onto conventional SRAM bitcells across the array.

At the bitcell level, the storage core is a standard 6T SRAM latch accessed by complementary bitlines (BL/BLB) and a wordline (WL). To support compute-in-memory, each cell includes a NOR compute gate and a cell multiplexer (MUX) that together conditionally couple the stored bit to the compute output. The compute gate realizes the per-cell interaction between the stored weight bit and the applied spin/control input (enabling or suppressing the contribution), so that only cells corresponding to active operands contribute to the column/row accumulation. The cell MUX provides fine-grained selection of the compute path, allowing the controller to time-multiplex bit-slices without disturbing the stored state. Collectively, the NOR gate-and-MUX augmentation enables the array to reuse dense SRAM storage for both coefficient retention and in-situ arithmetic, while keeping spin states and update control in peripheral digital logic (see Figure \ref{sram_array}).

\subsection{Annealing via Noisy Pseudo-Read}

\begin{figure}[t]
    \centering
    \includegraphics[width=0.45\textwidth]{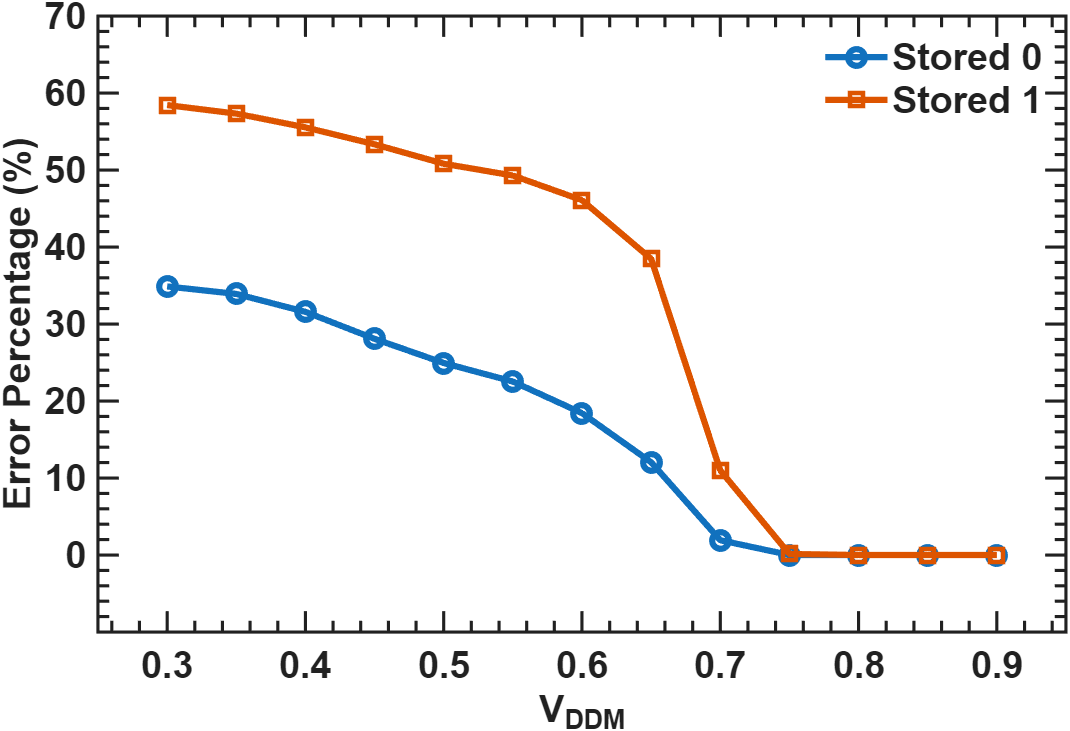}
    \caption{Measured pseudo-read error probability versus $V_{\mathrm{DDM}}$ for SRAM cells storing `0' and `1' from a taped-out DCIM prototype chip at TSMC 28nm process.}
    \label{assymetric error rate}
\end{figure}

To emulate the stochasticity required for annealing, a periodic pseudo-read operation is applied to the SRAM weight array while sweeping the memory supply $V_{\mathrm{DDM}}$ from a low value to a higher value. Reducing $V_{\mathrm{DDM}}$ intentionally weakens the SRAM static noise margin (SNM), so that the read access becomes increasingly non-ideal and induces probabilistic bit flips in the stored weight cells. The resulting perturbations manifest as an error rate that is experimentally characterized as a function of $V_{\mathrm{DDM}}$ and, critically, is asymmetric between stored `0' and `1' values (i.e., distinct $0\rightarrow1$ and $1\rightarrow0$ flip probabilities, see Figure \ref{assymetric error rate}). In this way, the architecture converts device-level variability into controlled, voltage-tunable temporal randomness, which serves as the annealing mechanism without requiring an explicit software temperature schedule or an external noise source.

To maintain a consistent signed representation during noisy operation, the pseudo-read perturbation is selectively applied only to the magnitude bits of each stored weight word, while the MSB that encodes the sign is held fixed, so that annealing-induced errors modulate weight magnitude but do not stochastically invert the sign. This constraint prevents large, non-physical jumps in the effective coupling matrix and preserves the intended polarity of interactions throughout the annealing sweep.

Following initialization, each annealing step proceeds by (i) applying a pseudo-read at the current $V_{\mathrm{DDM}}$ to realize the desired error profile in the effective weights, (ii) computing the local flip cost $\Delta E_i$ for a selected index under the current configuration, and (iii) updating the corresponding spin according to the sign of $\Delta E_i$ (with the stochasticity already embedded through the pseudo-read–corrupted weights). In contrast to a synchronous parallel rule, the control flow follows a sequential update: spins are visited one-by-one in a deterministic scan order (e.g., $i=1,2,\ldots,N$), and each decision is committed immediately before evaluating the next index. This removes the need for an on-chip random index generator, since exploration is already provided by the voltage-tunable pseudo-read perturbations in the stored weights, which randomize the effective $\Delta E_i$ encountered during the scan.

As $V_{\mathrm{DDM}}$ is swept from a lower value to a higher value, the pseudo-read error rate decreases: the early low-$V_{\mathrm{DDM}}$ portion of the schedule therefore injects stronger randomness that promotes exploration, whereas the later high-$V_{\mathrm{DDM}}$ regime becomes near-deterministic and primarily stabilizes the configuration as the system approaches convergence.

\subsection{In-Memory Hamiltonian Update Computation}

Equation~\eqref{eq:deltaE_qubo} makes explicit that $\Delta E_i$ contains both (i) an off-diagonal dot-product term, which is naturally produced by an in-memory MAC, and (ii) a diagonal contribution $Q_{ii}$, which is not produced by a pure off-diagonal MAC unless it is handled explicitly (e.g., by a separate add path, a dedicated diagonal storage path, or an algorithmic reformulation).

To make the update rule compatible with a direct MAC on a coupling matrix whose diagonal is constrained to zero, we apply a pinned-one embedding. Specifically, we introduce an auxiliary binary variable $q_p$ that is clamped to $1$ throughout the optimization, and construct an embedded matrix $\widetilde{\mathbf{Q}}$ such that (i) $\widetilde{Q}_{ii}=0$ for all problem-variable indices $i$, (ii) $\widetilde{Q}_{ij}=Q_{ij}$ for $i\neq j$ within the problem block, and (iii) the original diagonal is represented as couplings to the pinned variable, $\widetilde{Q}_{ip}=\widetilde{Q}_{pi}=Q_{ii}/2$. With $q_p\equiv 1$, this preserves the original objective up to a constant offset, while ensuring that the diagonal of the problem block is structurally zero.

Under this embedding, the per-variable flip cost can be evaluated using only the MAC-produced row sum
\begin{equation}
s_i \;=\; \sum_{j} \widetilde{Q}_{ij} q_j,
\end{equation}
leading to the diagonal-free update expression
\begin{equation}
\Delta E_i \;=\; 2\big(1-2q_i\big)\,s_i,
\label{eq:deltaE_qubo_pinned}
\end{equation}
where the contribution that previously appeared as $Q_{ii}$ in \eqref{eq:deltaE_qubo} is now included implicitly via the pinned coupling term inside $s_i$ (since $q_p=1$). In our flow, each iteration performs a \emph{sequential} update. Specifically, we visit variables one at a time according to a fixed scan order (or a randomized permutation) and compute $\Delta E_i$ for the current index $i$. If $\Delta E_i<0$, we flip the variable immediately, i.e., $q_i \leftarrow 1-q_i$; otherwise, $q_i$ is left unchanged. The energy change for subsequent indices is evaluated using the latest updated state.

Finally, to maintain a controlled annealing trajectory under voltage-scaled pseudo-read perturbations, the weight array is periodically restored (refreshed) to its nominal programmed values. This restoration resets accumulated pseudo-read-induced drift in the stored weights, while the reduced-$V_{\mathrm{DD}}$ operating point sets the instantaneous perturbation strength during each iteration; together, these mechanisms implement an annealing schedule in which flip decisions are made from the in-memory $\Delta E$ evaluation, and the underlying coupling matrix is re-established at a programmable cadence.

\subsection{Solution Tracking and Termination}

\begin{figure}[t]
    \centering
    \includegraphics[width=0.25\textwidth]{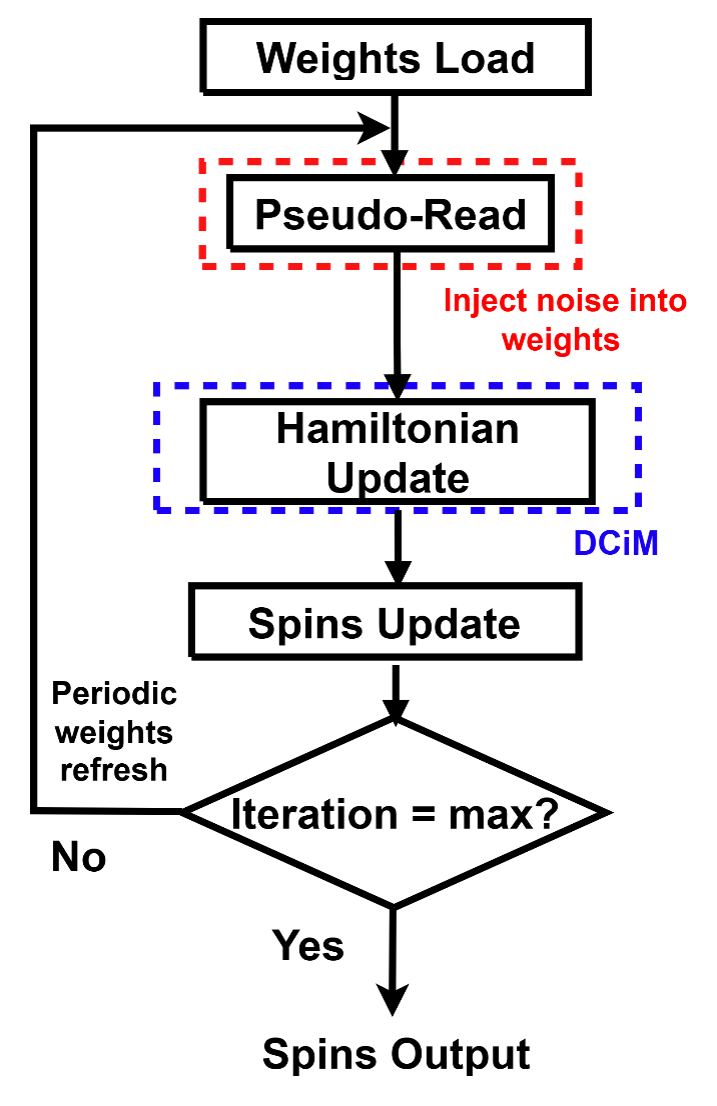}
    \caption{Control flow of the proposed DCIM-based annealer}
    \label{control flow}
\end{figure}

In~\cite{10.1145/3649329.3655673}, the control flow is formulated as an accept/reject dynamics over candidate configurations: the hardware evaluates the cost fucntion for the current configuration and a proposed updated configuration, compares them using local digital logic and registers, and then commits the update if it satisfies the acceptance condition. Accordingly, the reported solution is effectively the terminal spin configuration at the end of the annealing schedule, rather than an explicitly maintained global best-so-far state stored on-chip throughout the sweep. 

In our flow, we adopt the same philosophy: the solver reports the terminal state at the end of the annealing sweep, without maintaining an explicit global best-so-far configuration on-chip. Concretely, the DCIM array repeatedly evaluates the per-variable flip cost $\Delta E_i$ under the current state and commits updates according to the acceptance rule, while the annealing schedule is chosen to gradually suppress randomness and stabilize the configuration toward the end of the run. As a result, no additional storage or off-chip bookkeeping is required beyond the state registers already used for iterative updates, keeping the control and memory overhead minimal.

The overall annealing control flow is summarized in Figure \ref{control flow}, illustrating the sequence of weight loading, voltage-controlled pseudo-read noise injection, Hamiltonian update evaluation, spin update, and periodic weight refresh until termination.

\section{Results and Discussion}

We consider a feed-forward binary neural network (BNN) with $L$ layers which can be defined for a binary input vector $\mat{y}^{0}$ as:

\begin{equation}
    y = \mathrm{BNN}(\mat{x}) =
    \sgn\!\left(
        \mat{W}^{L-1}
        \sgn\!\left(
            \cdots
            \sgn(\mat{W}^{0}\mat{y}^{0})
            \cdots
        \right)
    \right).
\end{equation}
where $\mat{W}^{l}$ denotes the weight matrix associated with the layer $l$. The non linearity is introduced via $\sgn(\cdot)$, which is the element-wise sign function.

The BNNs are pre-trained on the MNIST~\cite{mnist} dataset using different input sizes, which are varied to control the number of variables in the final QUBO instance. Table~\ref{tab:BNN} summarizes the considered BNN configurations, including input size, layers, and the corresponding test accuracies. These models are not intended to match state-of-the-art BNN performance; rather, they serve as representative benchmarks to demonstrate that our DCIM Ising machine can identify adversarial perturbations even when the returned solutions are not fully constraint-satisfying (i.e., do not necessarily correspond to the global minimum of the constructed QUBO). These BNNs are then used to generate QUBOs for both classes which are reported in Table~\ref{tab:QUBO}. The dataset is cleaned to remove repeating and contradicting input-labels pairs, as we use adaptive averaging to reduce the input size for our formulation. 

\begin{table}[t]
\begin{center} % Add this
\begin{tabular}{|c|c|c|}
\hline
\textbf{Input Size} & \textbf{BNN Layers}   & \textbf{Accuracy} \\ \hline
% 5x5        & {[}31, 7, 1{]}   & 10\%     \\ \hline
7x7        & {[}63, 7, 1{]}   & 16\%     \\ \hline
11x11      & {[}127, 7, 1{]}  & 70\%    \\ \hline
28x28      & {[}1023, 7, 1{]} & 56\%    \\ \hline
\end{tabular}
\vspace{0.1cm}
\caption{Pre-trained BNNs with different input sizes and two class classification.}
\label{tab:BNN}
\end{center} 
\end{table}

\begin{table}[t]
\begin{center}
\begin{tabular}{|c|c|c|c|c|}
\hline
\textbf{BNN} & \textbf{Label} & \makecell{\textbf{Perturbation} \\ \textbf{Bounds}} & \textbf{Constraints} & \makecell{\textbf{Variables} \\ \textbf{(Spins)}} \\ \hline
63x7x1  & 0 & 16  & 4492  & 183 \\ \hline
63x7x1  & 1 & 16  & 4304  & 183 \\ \hline
127x7x1 & 0 & 32  & 15944  & 319 \\ \hline
127x7x1 & 1 & 32  & 19291  & 319 \\ \hline
1023x3x1 & 0 & 256 & 498700 & 1066 \\ \hline
1023x3x1 & 1 & 256 & 487771 & 1066 \\ \hline
\end{tabular}
\vspace{0.1cm}
\caption{Number of variables (spins) and constraints in the generated QUBO instances for the corresponding BNN configurations under their respective perturbation bounds.}
\label{tab:QUBO}
\end{center}
\end{table}

\subsection{Resultant Perturbations and Energy Profile}

A perfect solution corresponds to the global minimum of the constructed QUBO. In practice, however, identifying the global optimum is challenging due to the combinatorial nature of the problem and the highly rugged optimization landscape. Moreover, not all samples returned by an Ising machine (or a classical sampling based solver) are equally useful: even solutions that do not fully satisfy all constraints can still encode perturbations that flip the network prediction. Concretely, we extract the perturbation variables from a candidate good solution vector, reconstruct the perturbed input $\mathbf{x}'$, and then verify via BNN inference whether the predicted label changes.

We refer to solution vectors whose energies fall below a prescribed \emph{cutoff} as \emph{good solutions}. A perturbed sample $\mathbf{x}'$ is classified as a good solution if and only if
\begin{equation}
    \mathbf{x}' \in \{\text{Good Solutions}\}
    \iff
    \mathbf{x'}^{\mathrm{T}}\mathbf{Q}\mathbf{x'} \leq E_{\text{cutoff}},
    \label{Eq:GoodSol}
\end{equation}
where $E_{\text{cutoff}}$ denotes the energy threshold.

The cutoff is defined relative to the minimum observed energy. Let $E_{\min}$ denote the minimum energy calculated during QUBO formulation. We retain candidate solutions whose energies lie within the lowest 5\% energy band above $E_{\min}$, i.e., solutions whose objective values are within a 5\% relative gap of the minimum energy. 

Table~\ref{tab:results_cutoff} reports the number of good solutions identified for each QUBO instance using the proposed DCIM Ising machine and simulated annealing~\cite{Pardalos2009}. The results indicate that the proposed DCIM architecture consistently matches or surpasses simulated annealing in the number of near-optimal (cutoff-qualified) solutions recovered, while operating with lower energy consumption and reduced execution time.

\begin{table}[t]
\centering
\begin{tabular}{|c|c|c|c|}
\hline
\textbf{BNN} & \textbf{Label} & \textbf{DCIM Solver} & \textbf{Simulated Annealing} \\ \hline

63x7x1  & 0 & 2000 & 2000 \\ \hline
63x7x1  & 1 & 2000 & 1999  \\ \hline

127x7x1 & 0 & 2000 & 1989 \\ \hline
127x7x1 & 1 & 1999 & 1997 \\ \hline

1023x3x1 & 0 & 2000 & 1304 \\ \hline
1023x3x1 & 1 & 1998 & 1305  \\ \hline
\end{tabular}
\caption{Number of \emph{good} (cutoff-qualified) solutions obtained using the proposed DCIM Ising machine and simulated annealing for different BNN configurations.}
\label{tab:results_cutoff}
\end{table}

Solutions that lie near the global minimum (but are not globally optimal), referred to here as \emph{imperfect solutions}, may not satisfy all encoded constraints. However, this does not necessarily render them unusable. In the QUBO construction, certain constraints—such as auxiliary-variable consistency constraints within the BNN inference layers and most-significant-bit enforcement constraints—are less critical than the output-label mismatch constraint and the perturbation-budget constraint. Consequently, even partially constraint-violating solutions may encode valid adversarial perturbations.

To evaluate this effect, we extract perturbation vectors from all candidate solutions and reconstruct the corresponding perturbed inputs. These inputs are then verified through forward BNN inference. If the predicted label differs from that of the original input, the solution is counted as a successful adversarial attack, thereby certifying non-robustness at that sample.

Table~\ref{tab:results_succesful} reports the number of successful attacks obtained from our DCIM solver compared with simulated annealing solutions solutions for each QUBO instance. The results demonstrate that a substantial fraction of imperfect solutions still yield valid adversarial perturbations, highlighting the practical utility of near-optimal solutions in robustness verification.

\begin{table}[tbp]
\centering
\begin{tabular}{|c|c|c|c|}
\hline
\textbf{BNN} & \textbf{Label} & \textbf{DCIM Solver} & \textbf{Simulated Annealing} \\ \hline

63x7x1  & 0 & 994   & 526  \\ \hline
63x7x1  & 1 & 53    & 197  \\ \hline

127x7x1 & 0 & 106   & 291  \\ \hline
127x7x1 & 1 & 373   & 460   \\ \hline

1023x3x1 & 0 & 438  & 1110  \\ \hline
1023x3x1 & 1 & 1510 & 484 \\ \hline

\end{tabular}
\caption{Number of successful adversarial attacks obtained using the proposed DCIM Ising machine and simulated annealing for different BNN configurations.}
\label{tab:results_succesful}
\end{table}

The results reported above were obtained using higher than 8-bit precision, since the QUBO coefficients derived from the BNNs are non-uniform and span a dynamic range that exceeds native 8-bit representation. The smaller QUBO instances can be rescaled to operate within 8-bit precision hardware; however, any performance differences arising from this scaling are attributable to the quantization of the QUBO coefficients rather than to the Ising machine or its annealing dynamics.

\begin{figure}[htbp]
    \centering
    \includegraphics[width=\linewidth]{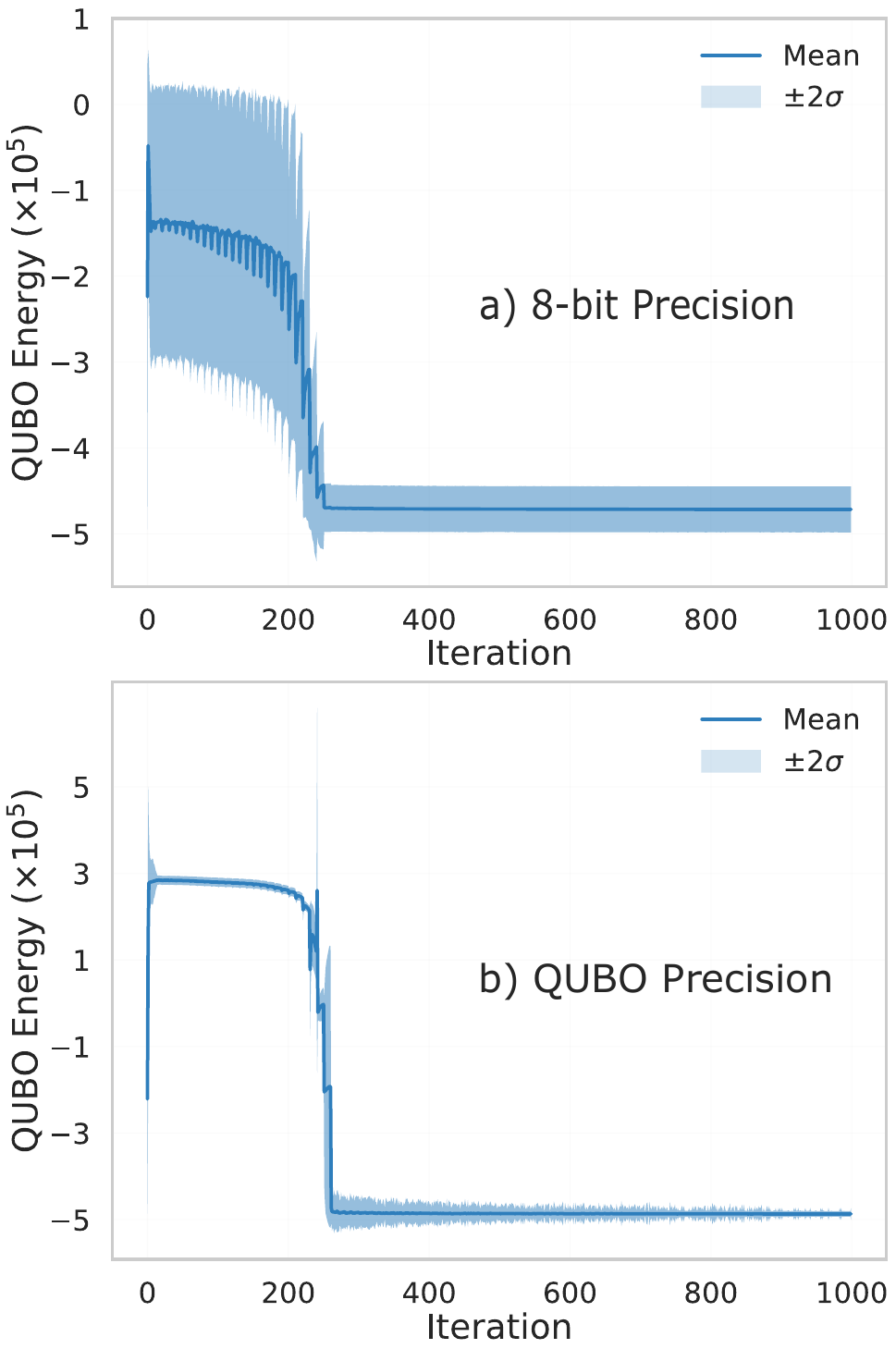}
    \caption{Energy evolution during annealing for the 1023x3x1 QUBO instance under (a) 8-bit quantized coefficients and (b) full QUBO-level precision.}
    \label{fig:EnergyPlot}
\end{figure}

As shown in Fig.~\ref{fig:EnergyPlot}, the 8-bit scaled QUBO converges to energy values that remain close to those obtained under full QUBO-level precision on average. Here, QUBO-level precision refers to the bit width required to represent the largest unsigned coefficient along with an additional sign bit. The comparison enables isolation of quantization effects from solver dynamics. Also, the observed increase in variance under 8-bit scaling can be attributed to coefficient quantization rather than to the intrinsic behavior of the Ising machine.

\begin{figure}[tbp]
    \centering
    \includegraphics[width=\linewidth]{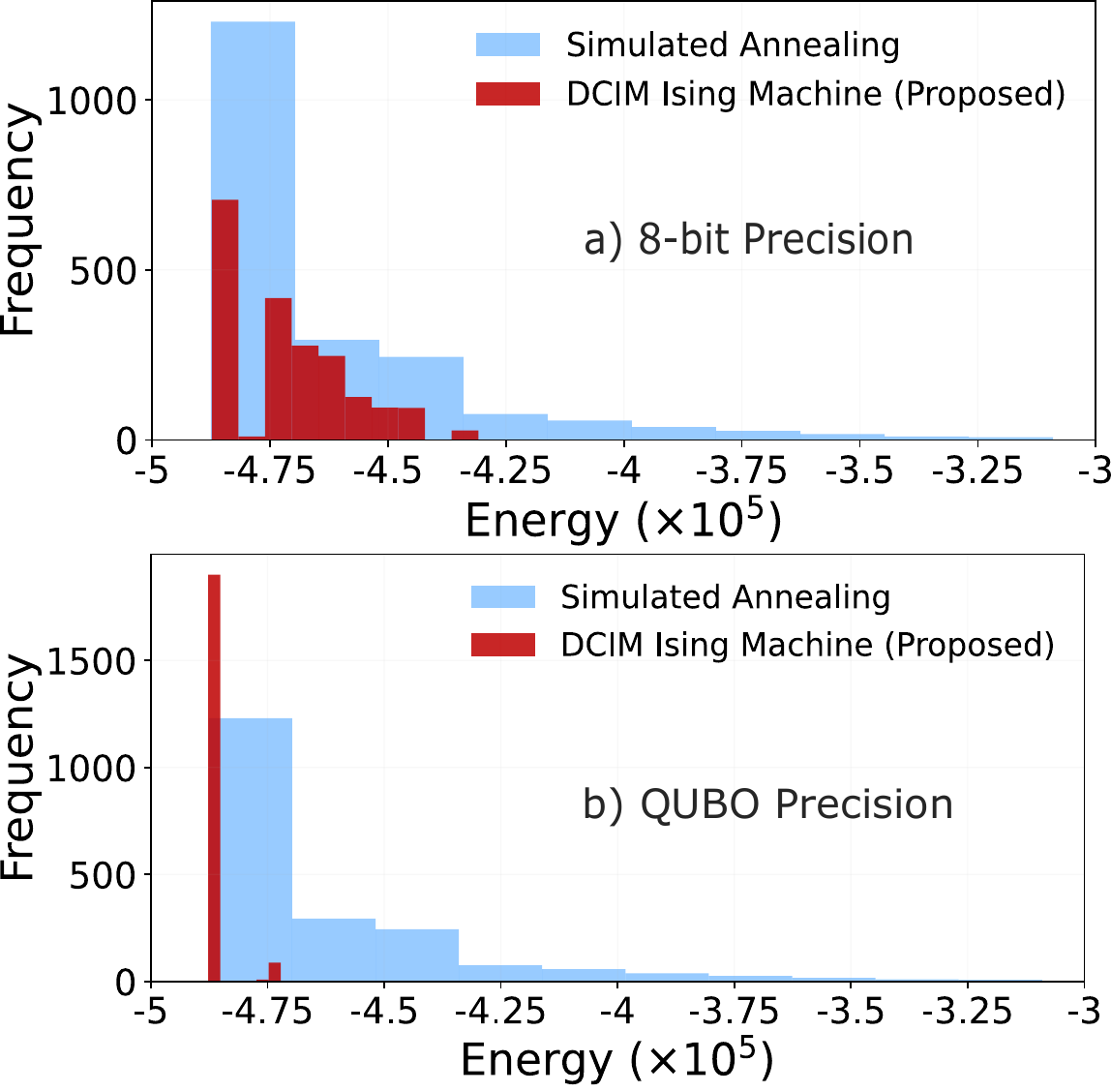}
    \caption{Energy distributions over 2000 samples comparing a) 8-bit quantized and b) full QUBO-level precision implementations of the proposed DCIM Ising machine.}
    \label{fig:EnergyHist}
\end{figure}

This trend is further illustrated in Fig.~\ref{fig:EnergyHist}, where the number of samples within the near-optimal energy band decreases under 8-bit scaling. Nevertheless, in both precision settings, the solver consistently produces solutions within the near-global-minimum region.

Another important aspect of the obtained solutions is whether the corresponding perturbations are \emph{unique}. A successful attack is counted as unique if its perturbation vector has not been observed previously, i.e.,
\begin{equation}
    \boldsymbol{\tau} \notin \mathcal{T}_{\mathrm{seen}},
    \label{Eq:Unique}
\end{equation}
where $\mathcal{T}_{\mathrm{seen}}$ denotes the set of previously recorded perturbation vectors.

\begin{figure}[bt]
    \centering
    \includegraphics[width=\linewidth]{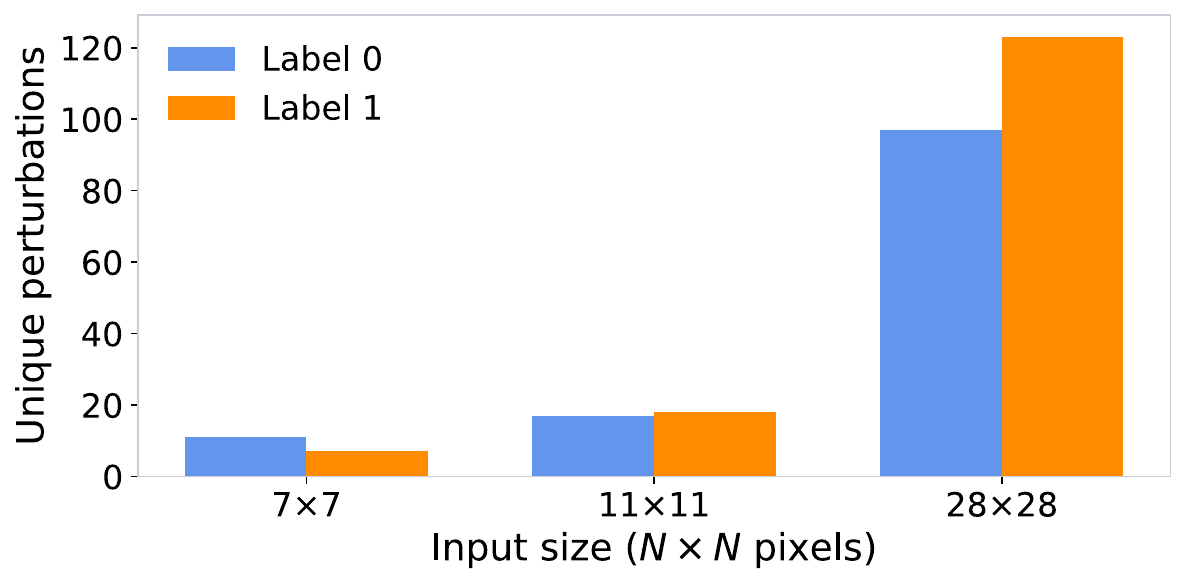}
    \caption{Number of unique perturbations found by our DCIM solver that resulted into a successful attack.}
    \label{fig:unique}
\end{figure}

Fig.~\ref{fig:unique} illustrates the number of unique adversarial perturbations extracted from our DCIM Ising Machine solutions. A larger number of unique perturbations indicates the propsed hardware’s ability to explore diverse regions of the adversarial search space. This diversity suggests that continued annealing or extended sampling may converge toward distinct globally optimal perturbations, rather than repeatedly rediscovering the same near-optimal configurations.

\begin{figure}[th]
\centering
\subfloat[7$\times$7, Label: 0\label{fig:perturbed_7x7_0}]{
    \includegraphics[width=0.45\columnwidth]{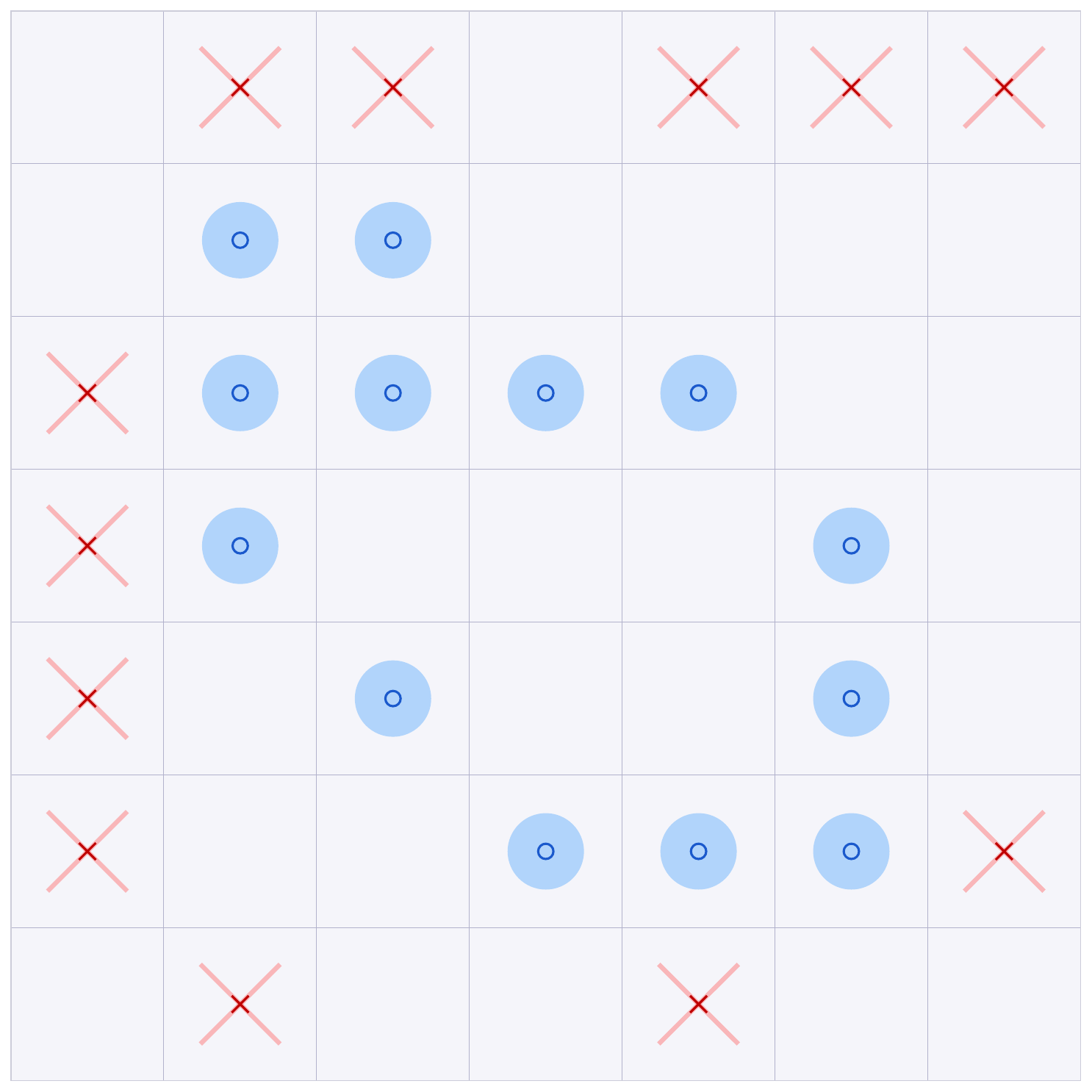}
}\hfill
\subfloat[7$\times$7, Label: 1\label{fig:perturbed_7x7_1}]{
    \includegraphics[width=0.45\columnwidth]{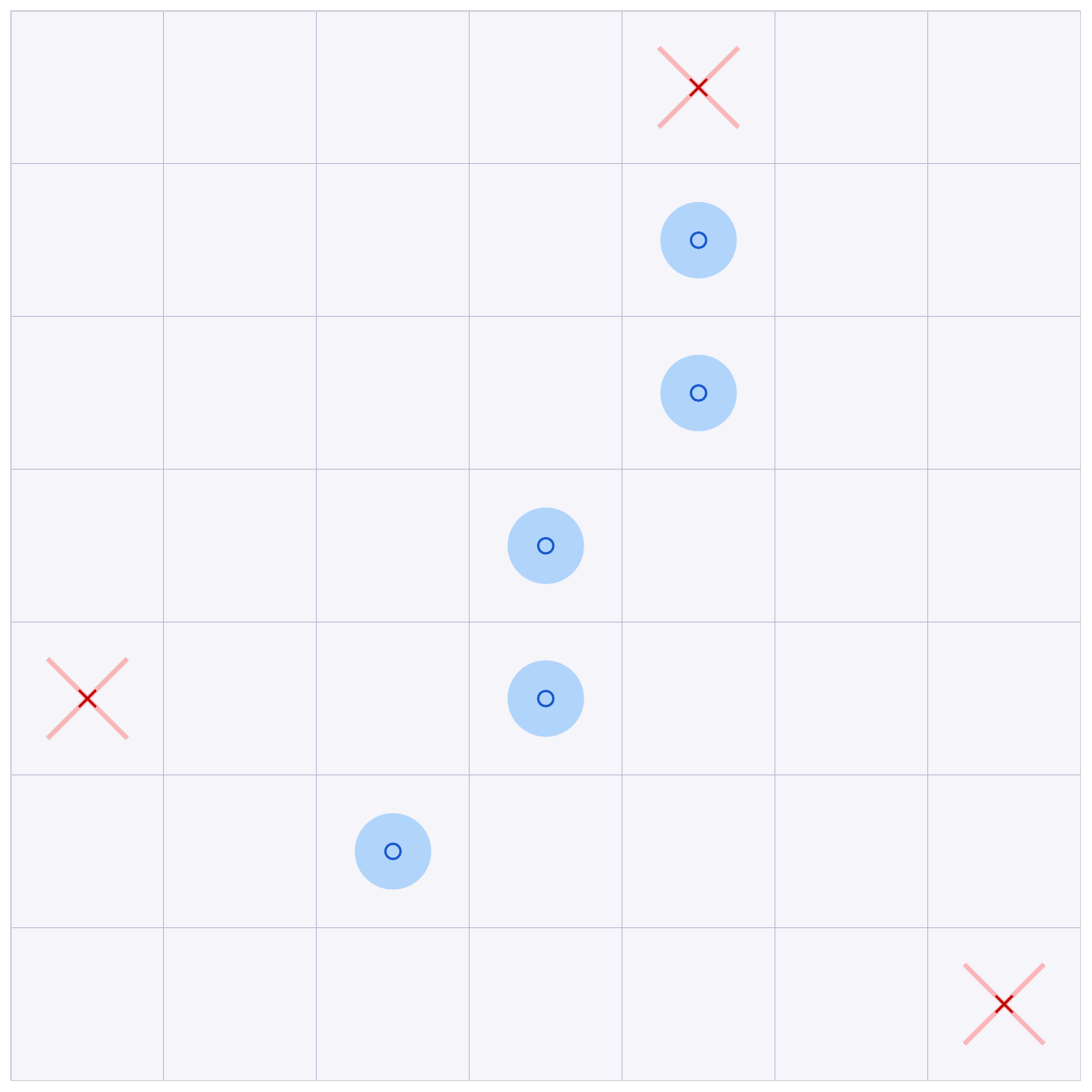}
}
\vspace{0.25cm}
\subfloat[11$\times$11, Label: 0\label{fig:perturbed_11x11_0}]{
    \includegraphics[width=0.45\columnwidth]{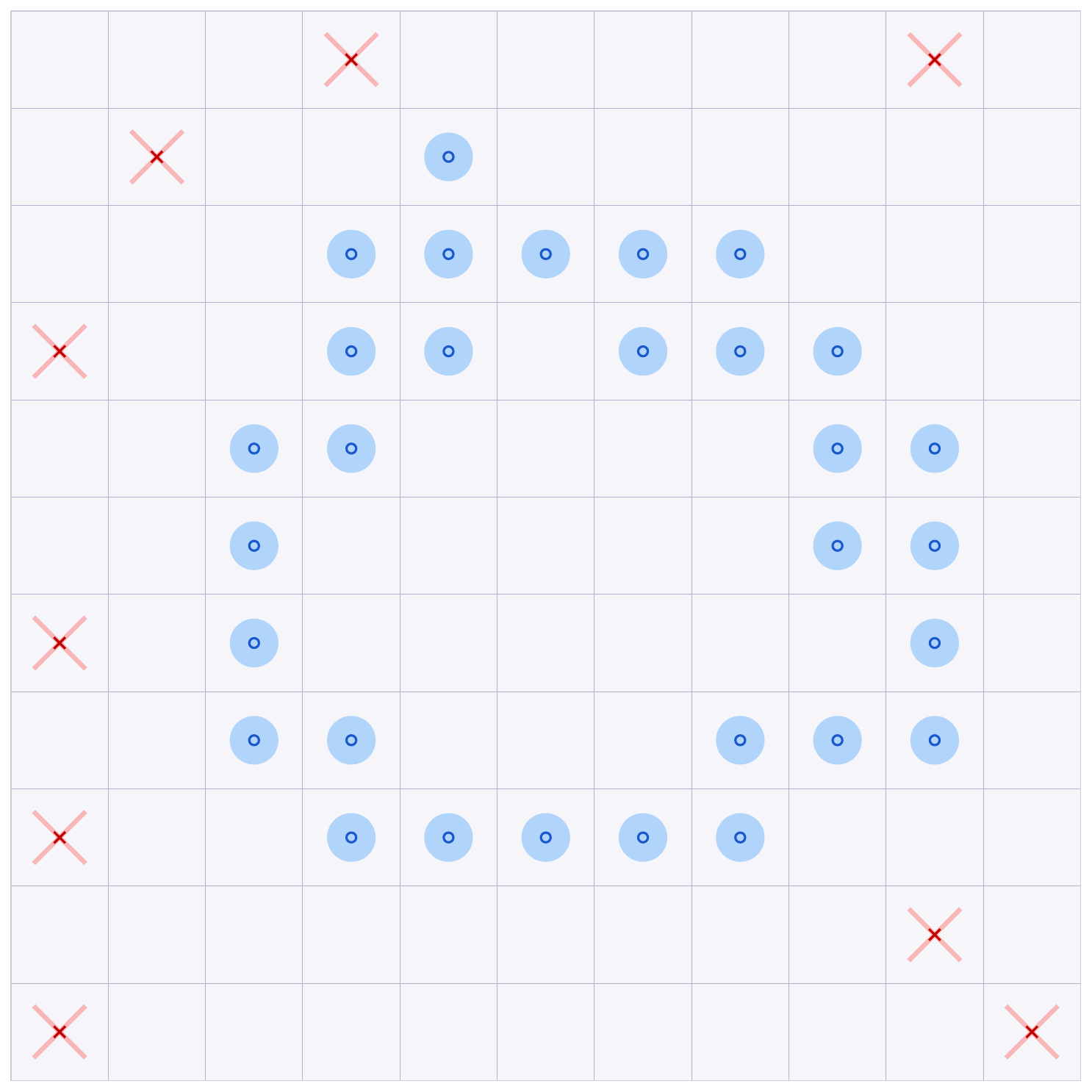}
}\hfill
\subfloat[11$\times$11, Label: 1\label{fig:perturbed_11x11_1}]{
    \includegraphics[width=0.45\columnwidth]{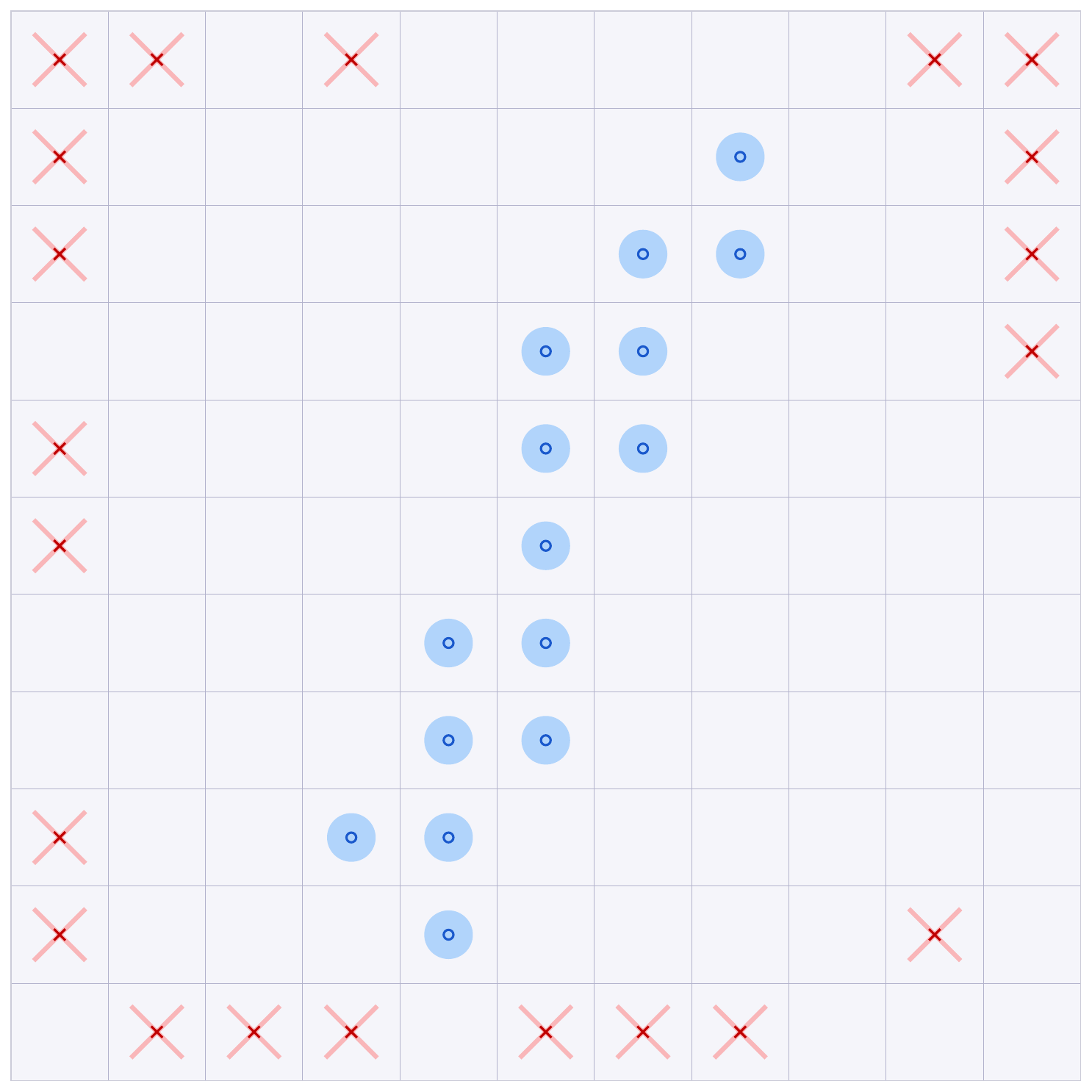}
}

\caption{Adversarial perturbations obtained from the proposed DCIM Ising machine for reduced-size BNNs ($7\times7$ and $11\times11$), demonstrating successful label flips for both classes.}
\label{fig:perturbed_examples}
\end{figure}

Fig.~\ref{fig:perturbed_examples} illustrates adversarially perturbed inputs obtained for smaller QUBO instances, which can be efficiently solved using 8-bit precision. The corresponding BNNs are trained for two-class classification (digits 0 and 1), and dimensionality reduction of the input images preserves sufficient discriminative structure to maintain class separability. Consequently, reduced-resolution inputs remain adequate for demonstrating adversarial vulnerability in this binary setting.

\begin{figure}[th]
\centering

\subfloat[Label 0\label{fig:perturbed_28x28_0}]{
    \includegraphics[width=0.45\columnwidth]{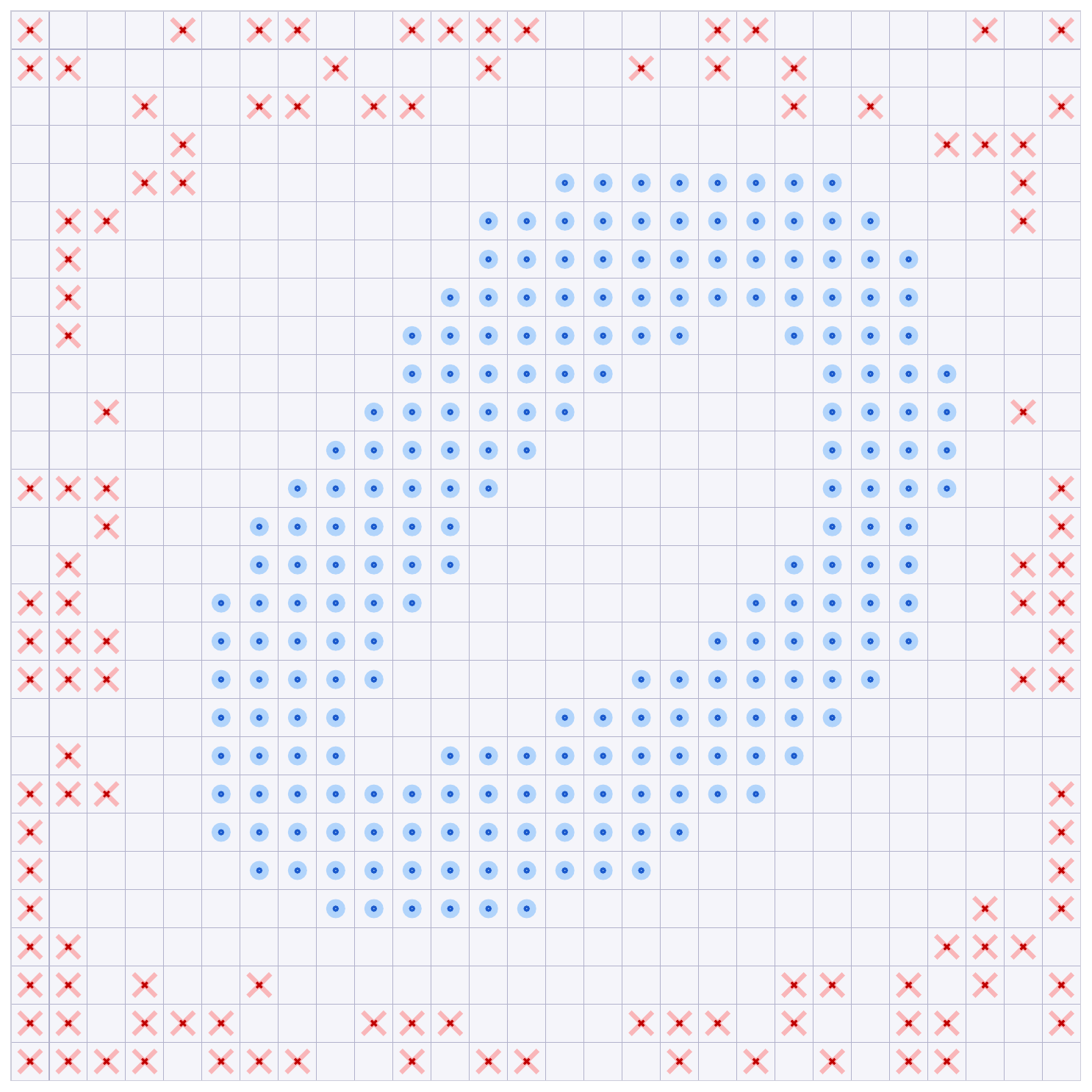}
}\hfill
\subfloat[Label 1\label{fig:perturbed_28x28_1}]{
    \includegraphics[width=0.45\columnwidth]{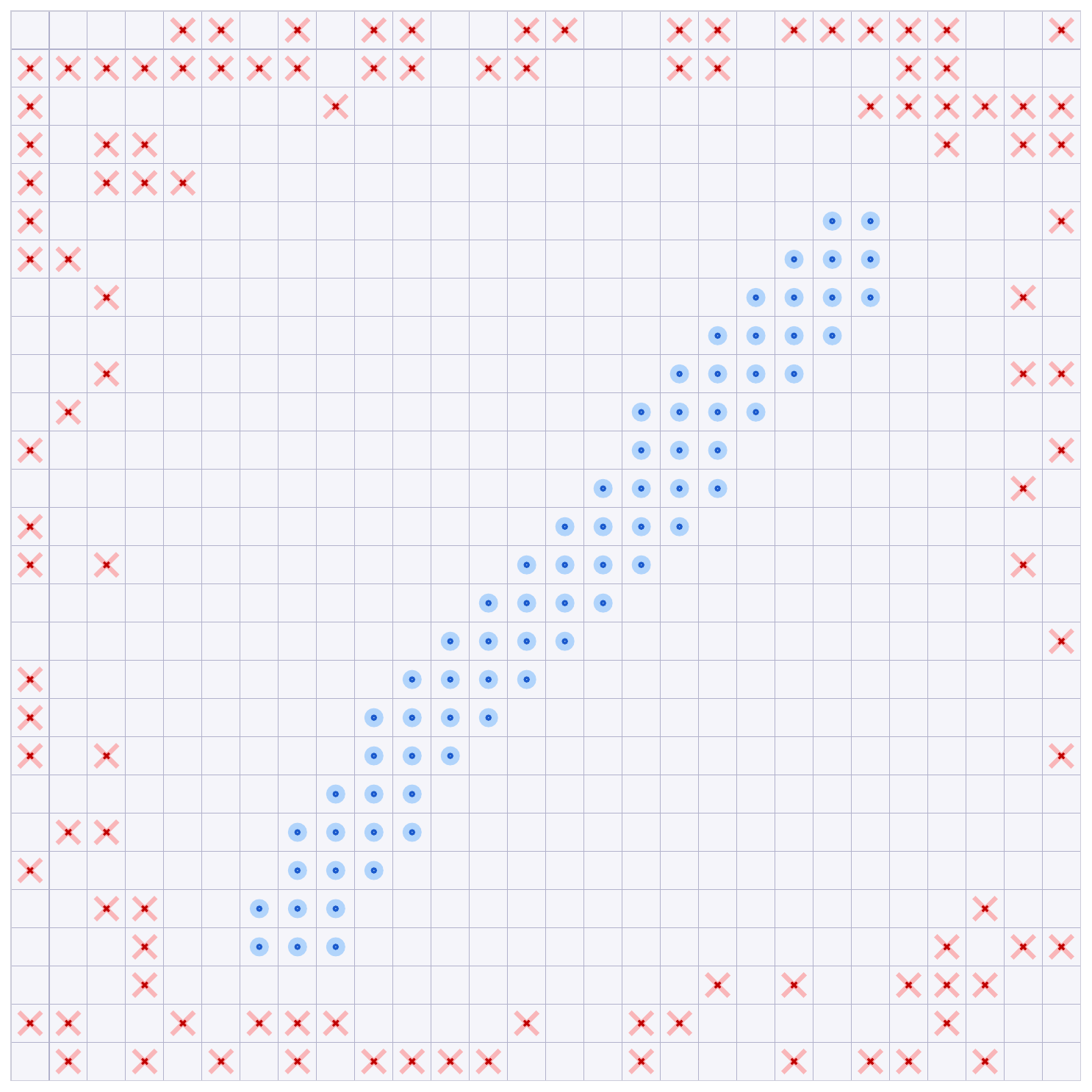}
}

\caption{Adversarial examples generated for the full-resolution ($28\times28$) BNN, confirming that our proposed DCIM Ising Machine verification approach scales to high-dimensional inputs.}
\label{fig:perturbed_28x28}
\end{figure}

However, for classification tasks involving a comparitively higher number of classes, full-resolution inputs are generally required to preserve all discriminative features. As shown in Fig.~\ref{fig:perturbed_28x28}, adversarial perturbations were successfully generated for the full-size ($28\times 28$) BNN as well, confirming that the proposed approach scales to higher-dimensional instances.

\subsection{Hardware PPA and Benchmarking}

To estimate the hardware PPA of the proposed design, we adopt a first-order projection based on the SRAM DCIM prototype reported in~\cite{Kong2026Compact}. This choice is motivated by the close architectural correspondence between the two systems, including SRAM-based in-memory storage of quantized weights, DCIM-style accumulation for update evaluation, and a controller-driven iterative annealing flow. Accordingly, we scale the reported area and timing metrics in~\cite{Kong2026Compact} to our target problem dimensions and precision to obtain a conservative PPA estimate for our implementation.

The largest instance considered in Table \ref{tab:QUBO} is a $1066\times 1066$ QUBO, which we embed into a $1067\times 1067$ weight matrix (including the additional pinned node), hence requiring $1067^2$ matrix entries to be stored in the DCIM SRAM array. With 8-bit quantization implemented, the total number of stored SRAM bits is $1067^2\times 8 = 9{,}112{,}712$ bits ($\approx 9.11$ Mb).
From the SRAM DCiM TSP chip demonstrated in~\cite{Kong2026Compact}, the reported DCiM bitcell footprint is approximately $0.7~\mu\mathrm{m} \times 2.43~\mu\mathrm{m}$, i.e., $A \approx 1.701~\mu\mathrm{m}^2$ per stored bit. For a weight array requiring approximately $9.11$~Mb of SRAM storage, the corresponding array area is estimated as $9.11\times 10^6 \times 1.701~\mu\mathrm{m}^2 \approx 1.55\times 10^7~\mu\mathrm{m}^2 \approx 15.5~\mathrm{mm}^2$ (weights only). Assuming the same area efficiency reported in~\cite{Kong2026Compact}, where the DCIM bitcell array occupies $48.7\%$ of the total macro area, the total area of our design can be projected by scaling the estimated weight-array area accordingly. Using the weights-only estimate of $15.5~\mathrm{mm}^2$, the corresponding total area is $31.8~\mathrm{mm}^2$.

In our simulated annealing baseline, executed on an AMD Ryzen\texttrademark{} Threadripper\texttrademark{} workstation-class processor~\cite{AMDThreadripperWorkstations}, 1000 Monte Carlo sweeps over 1066 spins for single sweep require approximately 20.29~s of wall-clock time and 1660.37~J of CPU package energy. This corresponds to an average CPU package power of 81.83~W. Each update involves a length-$N$ vector interaction with the dense $Q$ matrix, resulting in a memory-bound computation pattern characteristic of von Neumann architectures, where data movement between memory and compute units dominates runtime.

Under the same clock frequency and operating assumptions as the SRAM DCIM chip in~\cite{Kong2026Compact}, the projected power consumption and time-to-solution of our design are estimated to be $53.194~\mathrm{mW}$ and $113.85~\mathrm{ms}$, respectively. In contrast to the software baseline, our SRAM-based DCiM Ising machine performs the required matrix--vector interactions directly within memory, enabling highly parallel spin-interaction evaluation without repeated data movement between memory and compute units. This architectural shift 
eliminates the energy-intensive data movement between memory and compute units, 
yielding an estimated $178\times$ acceleration in convergence rate and a 
$1538\times$ improvement in power efficiency relative to conventional 
CPU-based implementations.

\section{Conclusion}

Recent work has shown that the robustness verification of binary neural networks (BNNs) can be formulated as a QUBO instance whose solution certifies whether an adversarial perturbation exists within a given perturbation budget. However, the resulting QUBO instances are typically large and characterized by highly nonconvex energy landscapes, making the search for globally optimal solutions computationally demanding when executed on conventional solvers.

In this work, we have presented an alternative method that leverages imperfect solutions obtained from an SRAM-based DCIM Ising machine. Instead of requiring perfect solutions that satisfy all encoded constraints, our approach exploits the fact that imperfect solutions can already contain valid adversarial perturbations capable of revealing the non-robustness of the BNN. This observation enables a practical verification paradigm in which near-optimal states of the QUBO are sufficient to discover adversarial examples without explicitly solving the optimization problem to global optimality.

We have demonstrated the proposed workflow on BNNs of varying input dimensions derived from the MNIST dataset. Experimental results show that the SRAM-based DCIM Ising architecture consistently identifies near-global-minimum solutions and produces a large number of successful and unique adversarial perturbations. Comparative analysis with simulated annealing further indicates that the proposed hardware achieves comparable or superior performance while maintaining favorable energy–time characteristics.

Overall, this work establishes an end-to-end framework that combines QUBO-based BNN robustness formulations with Ising-based hardware optimization. By exploiting imperfect solutions for verification, the proposed approach provides a scalable pathway for applying unconventional computing platforms, such as Ising machines, to robustness analysis and trustworthy artificial intelligence systems.

\section*{Data Availability}
The code used to generate the QUBO instances for two-class classification binary neural networks is available at: \url{https://github.com/Rahps97/Ising-Machine-BNN-Verification.git}.

\section*{Acknowledgments}
AI Chatbot tools were used to assist with grammar refinement; all scientific content and analysis were created solely by the authors. The authors would like to thank Dr. Seyran Saeedi of UCSB for initial formulation of the problem. This work was supported in part by NSF Grant 2311295, and by Intel Emerging Technology Strategic Research Sector (SRS) funding and mentored by Dr. Hai Li and Dr. Ian Young, Exploratory Integrated Circuits, Components Research, Intel Corporation.

\vspace{11pt}
\bibliographystyle{IEEEtran}
\bibliography{ref.bib}

\end{document}